\documentclass[]{interact}
\usepackage{lineno,hyperref,pifont,graphicx,xcolor,amsmath,lscape,multirow,longtable, algorithm, algorithmic, amssymb}
\graphicspath{{./}}
\usepackage{epstopdf}
\usepackage[caption=false]{subfig}

\theoremstyle{plain}

\theoremstyle{definition}

\theoremstyle{remark}

\begin{document}

\articletype{Research Article}

\title{A New Intrusion Detection System using the Improved Dendritic Cell Algorithm}

\author{
\name{Ehsan Farzadnia* \textsuperscript{a}\thanks{CONTACT Ehsan Farzadnia. Email: e405504@gmail.com}, Hossein Shirazi \textsuperscript{a}\thanks{CONTACT Hossein Shirazi. Email: Shirazi@mut.ac.ir}, Alireza Nowroozi \textsuperscript{b}\thanks{CONTACT Alireza Nowroozi. Email: Alirezanowroozi@iribu.ac.ir}}
\affil{\textsuperscript{a} Faculty of Electronics and Computer, Malek-Ashtar University of Technology, IRAN; \\       
\textsuperscript{b} Department of Media Engineering, IRIB University, Tehran, IRAN}
}

\maketitle

\begin{abstract}
The Dendritic Cell Algorithm (DCA) as one of the emerging evolutionary algorithms is based on the behavior of the specific immune agents; known as Dendritic Cells (DCs). DCA has several potentially beneficial features for binary classification problems. In this paper, we aim at providing a new version of this immune-inspired mechanism acts as a semi-supervised classifier which can be a defensive shield in network intrusion detection problem. Till now, no strategy or idea has already been adopted on the $Get_{Antigen()}$ function on detection phase, but randomly sampling entails the DCA to provide undesirable results in several cycles in each time. This leads to uncertainty. Whereas it must be accomplished by biological behaviors of DCs in tissues, we have proposed a novel strategy which exactly acts based on its immunological functionalities of dendritic cells. The proposed mechanism focuses on two items: first, to obviate the challenge of needing to have a preordered antigen set for computing danger signal, and the second, to provide a novel immune-inspired idea in order to non-random data sampling. A variable functional migration threshold is also computed cycle by cycle that shows necessity of the Migration threshold (MT) flexibility. A significant criterion so called capability of intrusion detection (CID) used for tests. All of the tests have been performed in a new benchmark dataset named UNSW-NB15. Experimental consequences demonstrate that the present schema dominates the standard DCA and has higher CID in comparison with other approaches found in literature.
\end{abstract}


\section{Introduction}
\label{intro}
Evolutionary Computing (EC) as a subfield of artificial intelligence, can be defined as a representative form of a family of algorithms inspired by the theory of biological evolution. Among the EC schemas, a number of works have been presented based on meta-heuristic algorithms like Partial Swarm Optimization (PSO), Ant Colony Optimization (ACO), Krill Herd Algorithm (KHA) \cite{Ref55,Ref56,Ref57,Ref58,Ref68,Ref70} and Black Hole Algorithm(BHA) \cite{Ref69} which can be mentioned. Ref.\cite{Ref64,Ref65,Ref66,Ref67} are including works have focused on the Artificial immune system (AIS) field which is hot research topic for those issues where security is critical. AIS has most capabilities and give us required motivation to use its methods in some complex problems. It is defined as a computational paradigm that inspired by theoretical immune based principles and techniques. Capability of pattern recognition of immune system is used to distinguish between foreign cells entering the body, known as non-self-elements, and the body own cells, known as the self-elements. This process called “self and non-self-theory”. It is noted that Immune inspired methods cannot produce the same high performance as the human immune system. Hence, a newer immunological, named the Danger Theory (DT) as 2nd generation of AIS has been proposed. The DT is based on behavior of specific immune cells called Dendritic cells (DCs). It clarifies that the immune system looks for danger producing elements and events instead of discriminating between self and non-self. So, an inspiration from the behavior of DCs caused to development of a new bio-inspired algorithm named Dendritic Cell Algorithm (DCA). The DCA was successfully applied to a wide range of applications in literature. 
\\
Many researchers have caught their attention to investigate behavior of DCs so as to led to a variety of DCA versions aiming at improving the standard version of the DCA \cite{Ref15,Ref16,Ref17,Ref28,Ref32}.  Hence, this area is hot for research. In this paper, we have focused on building dynamism for the migration threshold of DCs and so propose a method for good sampling antigens by DCs to finally generate a novel semi-supervised classifier. Also, the problem of requirement for making antigen set in form of preordered to process Danger Signals has been solved. We followed two main goals in this paper: First one is to utilize the CSM signal to make MT variable cycle by cycle so as to provide a primary formula. Second one is to adopt a novel strategy for radius of presenting by DCs to sample multiplied antigens non-randomly. \\ 
The rest of the paper is organized as follows: Section 2 reviews literatures. In the next section, preliminaries about the methods which we basically worked on have been introduced. Section 4, presents our proposed system. Analyzing the effectiveness of the proposed schema has been verified in Section 5 through experiments. Last section draws a summary of this research paper and propose future works.
\\
\section{Related works}
In recent proceedings of research studies, using AIS methods, especially the dendritic cell algorithm (DCA), includes intrinsic capabilities so as to be a solution for anomaly detection problems \cite{Ref3,Ref15,Ref12}.  Various versions of the DCA has already been presented in some works that each one has different pros and cons depending on the type of problem \cite{Ref16,Ref17,Ref19}.  Some investigations show that the DCA has potential properties so as to be utilized for binary classification problems. In Ref.\cite{Ref16}, Chelly and Elouedi deeply study on summarizing the powerful characteristics of the DCA and its applications. In this work, we see some advantages: first, almost all of approaches for the DCA improvements and its limitations are evaluated and addressed well, and second, it is good to utilize as a baseline for proposed hybrid methods. The classical DCA has in excess of 10 parameters that employs many stochastic elements making it difficult to analysis. So, deterministic version of the DCA (dDCA) \cite{Ref17} is another option is formulated to help gain better insight into the underlying mechanism of the DC. The dDCA, consists of a controllable number of parameters. Main disadvantage of that is its failure on elevating deterministic performance at conditions that time slices between attacks tend to be low. On the other hand, the dDCA shouldn’t be used in networks with highly attacks. In this work, we used UNSW-NB15 dataset that is much better than KDDCup99. This benchmark dataset consists of a various range of attacks and normal traffics that rate of attacks to normal is balance.
\\    	  
Santanelli and et al in Ref.\cite{Ref15}, work on problematic issue of input signals assessment on the DCA. They optimize the probabilities of safe and danger signals using the GA. The goal is to calculate $p_{safe}$ and $p_{danger}$ values for the DCA that is their advantage. The most important drawback is that the dataset is outdated. In Ref.\cite{Ref33} which is a dissertation, authors investigate all of the key features of the DCA from two aspects of theoretical and practical for its development. In this work, a base foundation for investigating the signals calculation problem and preprocessing phase has been introduced. This work gives us a good view for generating input signals manually through dataset attributes which known as Knowledge of Experts (KoE). They also study KoE to attempt to provide new functions by analyzing features of NSL-KDD so as to find out three input signals, though stability on effectiveness of KoE is not approved. For an instance, permutation of three values of $count$, $src_{bytes}$ and $dst_{bytes}$ from KDDCup99 can be set to three input signals: safe signal (SS), PAMP signal and danger signal (DS). It can help us to inspire to provide ideas in signal assessment phase of the DCA. In our work, we have used IG method for signal generation from dataset. Our default approach to process three input signals (PAMP, SS and DS) is based on two mean or median criteria were also been analyzed. 
\\
A Cloud based algorithm presented for intrusion detection in Ref.\cite{Ref32} inspires of decentralized behavior of dendritic cells to implement the DCA on cloud. The goal is to make a decentralized mechanism for DC so as to model its activities completely similar to DCs activities. The important advantage of this work is that cloud environment can be very beneficial for developing DCA basis functionalities which is a challenge due to the demands of huge memory capacity. In Ref. \cite{Ref44}, author focuses on dendritic cell algorithm limitation when coping with very large datasets. She proposes a distributed version of the DCA based on MapReduced framework.
In Ref.\cite{Ref75}, an efficient distributed DCA is proposed based on partitioning the DCA processes into elementary tasks, and tested with 360 million antigens. In Ref.\cite{Ref74}, the challenge of the context assessment phase is obviated by integrating the K-means algorithm to the DCA in order to map the DCA cumulative semi-mature and mature context values into two clusters. Although used dataset (KDDCup99) is very old, the proposed approach is competitive compared with other commonly used classifiers. However, its performance is not guaranteed especially when traffic is online and change rapidly overtime. 
\\ 
Ref.\cite{Ref73} is very significant because it focuses on input signal generation challenge. It utilizes fuzzy logic to map from the input space to the output space. In this work, KDDCup99 is applied for input signal generation. Selected features for each DCA input signal (input space) are as follows: count and srv count for Danger Signal, logged in, srv different host rate and dst host count for Safe Signal and serror rate, srv serror rate, same srv rate, dst host serror and dst host rerror rate for PAMP.  This article gives us a good instance of how to get input signals from originally dataset. In work \cite{Ref46}, they work on optimizing weight function of the DCA that is a problem. They use genetic algorithm to generate optimized set of weights. For knowing how the DCA works, Ref.\cite{Ref17} as a prerequisite can be very beneficial. Regarding literature, neither of works introduce ideas to do preordering of antigen set, or to work on non-random data sampling. So, we addressed it well in our work. In the proposed IDS, dataset as antigen set is firstly divided into some sections known as the cell tissues based on two nominal attributes of protocol-service. In each cell tissue, each DC presents antigens (network traffic records) which have overlapping with them into its coverage. Each DC may cover one or more antigens and vice versa.
\\
\section{Preliminaries}
\subsection{\label{2.1}Dendritic cell algorithm} 
\begin{figure}[h]
\centering
\includegraphics[scale=0.5]{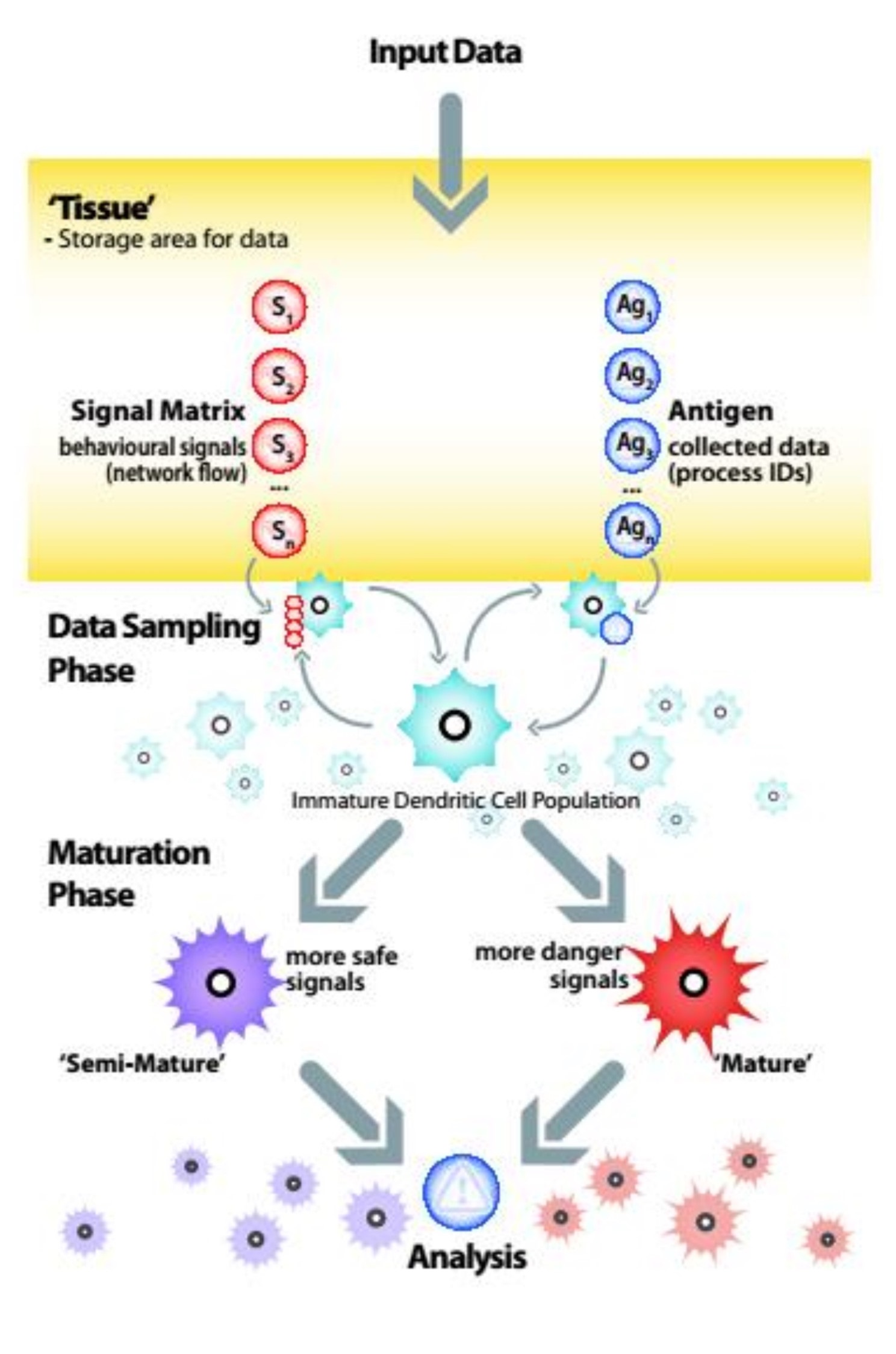}
\caption{\label{Fig.1} Dendritic cells migration procedure}
\end{figure}

\begin{figure}[h]
\centering
\includegraphics[scale=0.3]{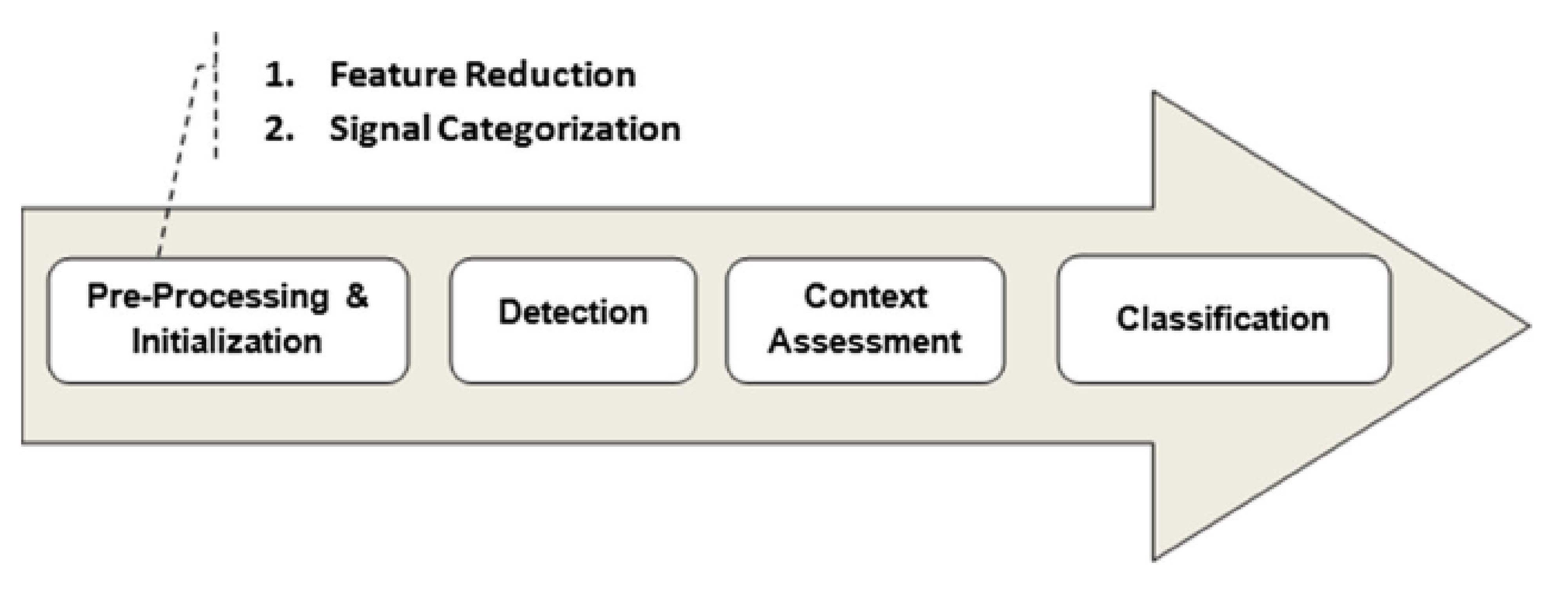}
\caption{\label{Fig.2} The DCA process, taken by Ref.\cite{Ref16}}
\end{figure}

The Danger Theory (DT) as the latest immunological theory firstly proposed by Matzinger in 1994 \cite{Ref15}.  The DT is an inspiration from the behavior of special immune cells inside body called the dendritic cells (DCs). Mediators of various immune responses, they are actually rapid messengers between innate and adaptive IS. Similar to NSA, the DT is also capable of discriminating between self-and non-self, but the main difference is that it looks for danger-producing elements and circumstances by assessing signals reflected from healthy cell tissues (safe) or injured ones. (Fig.\ref{Fig.1} ) Thus, the DT declares distinguishing antigens as “safe” and “non-safe” by IS while also as “self” and “non-self”.
\\
\\
The DCA is an abstract model of behavior of DCs. It is a kind of AIS that is based on the DT concept. Its process includes four phases: preprocessing, detection, context assessment, classification (Fig.\ref{Fig.2}).  Alg.4 at appendix also shows its original pseudo code. But, how does it work?

In body organism, each cell informs its safety status to nearest dendritic cells by signaling. They are absolutely like guardians that their task is to oversee tissues that means checkup \footnote{In computational perspective, checking up is accomplished by four functions: $Get_{antigens}$, $Get_{signals}$, $Calculate_{inter}$, $Update_{cumul}$}, not surgery. When an antigen  ($Ag_{i}$) breaches to cell tissues by skin, it probably harms them that its outcome leads to destruction. Injured and dead cells send danger/PAMP signals to immature DCs for assessment. This process called “signal assessment” is one of serious phases of the DCA. Signals sent from healthy/injured cells (considered as suspicious antigens) are received by some of DCs ({$DC_{j}={DC_{1},DC_{2},...,DC_{m}}$}). On the other hand, $n$ antigens with signals from all categories (SS, DS, and PAMP \footnote{Respectively safe, danger and PAMP signals}) must be presented by population of $m$ dendritic cells \footnote{That is a many to many relations} (distributed data sampling phase). The process of calculating input signals has been described in Ref. \cite{Ref16} in details. Calculating the cumulative output signals is also accomplished by presenter DCs.  When pervaded  $Ag_{i}$  is sampled by a subset of  DCs, they are stimulated into rest which have no chance for monitoring it.  These stimulations lead to the assignment of cell-contexts to DCs. At the end, immature $DC_{j}$  will have two cumulative values include semi-mature and mature in addition to cumulative CSM value that cause $DC_{j}$ to be able to stop monitoring and migrate if CSM exceeds MT; Otherwise, immature  $DC_{j}$   continues sampling antigens. Migrated DCs go to activate lymphocytes so that captured antigens are operated so better than past.  
\\
\\
According to Fig.\ref{Fig.2}, the preprocessing phase involves two primary steps: At first, the most important attributes from the gold data (train set) are selected and then, each attribute is mapped to an appropriate signal category. An immunological definition for input signals has been stated by Ref.\cite{Ref16}.   Two current methods to assess input signals are feature selection and knowledge of experts (KoE). Both of them have key challenges. Dimensional reduction deals with selecting the best optimum subset among   $2^{n}$   possible subsets of features that is an NP-hard problem, and success of the second technique also relies on the traffic flow and how to be analyzed by information security scientists so as to find most significant features to map to signals categorization. So, we can say that success of the DCA has a direct relationship with obtained signals, and this is an open problem. The weighted sum equation illustrates how the input signals are processed. It is defined as follows: (Taken by Ref.\cite{Ref16}):
\\
\begin{equation}\label{m1}
\footnotesize\ C = \frac{[(W_{PAMP}\times\sum_{i}{PAMP_{i}})+(W_{SS}\times\sum_{i}{SS_{i}})+(W_{DS}\times\sum_{i}{DS_{i}})]}{(W_{PAMP}+W_{SS}+W_{DS})}\times\frac{1+Inf}{2}
\end{equation}
\\
Where, $C_{|CSM,smDC,mDC|}$ , Inf is the inflammation signal which is assumed “1”, and weights are also gotten by an expert. Actually, each DC samples  $i$  antigens. Main weights matrix supposed by main work which has been made by Greensmith. Ref.\cite{Ref28} are used in research. It is defined as below:
\\
\begin{equation}\label{mtrx}
\footnotesize\ Weight\,\,coefficient\,\,matrix\,\,(WCM) =f 
\begin{vmatrix} Signals & PAMP& SS & DS \\ 
		         CSM & 2 & 1 & 2 \\
		         smDC & 0 & 0 & 2 \\
		         mDC & 2 & 1 & -2 \end{vmatrix}
\end{equation}
\\
The DC cumulative output signals (CSM, smDC, mDC) perform two roles: first, to decide what is an antigen type, and to suppress sampling the data. Furthermore, the MCAV is calculated by dividing the number of DCs in mature context that  $Ag_{i}$   has caused migrating them, by the total number of presented DCs have sampled to  $Ag_{i}$   when no immature DC didn’t remain in population.
\\
\begin{equation}\label{MCV}
\footnotesize\ MCAV=\frac{No.\,\,of\,\,mature\,\,DCs}{Total\,\,No.\,\,of\,\,DCs}
\end{equation}
\\
The degree of anomaly of a given antigen is assessed by the MCAV. Antigens MCAV values are compared with a determined anomalous threshold (at). It is defined as below:
\\
\begin{equation}\label{at}
\footnotesize\ at=\frac{No.\,\,of\,\,anomalious\,\,data\,\,items\,\,(an)}{Total\,\,No.\,\,of\,\,data\,\,items (tn)}
\end{equation}
\\
Those antigens have MCAV values greater than this threshold, and are classified into anomalous category while the others will be normal. The determination of this can be obtained by calculating reflection of the potential danger rate in distribution of incoming data between normal class and abnormal one. Besides, the MCAV values are discrete; 0 or 1, so it creates problem in our hybridization idea. In next sections, we have observed that some of these above equations do not have minimum proficiency to be used to contiguous problems like “network anomaly detection”.
\\
\section{Proposed system}
\begin{figure}[h]
\centering
\includegraphics[scale=0.35]{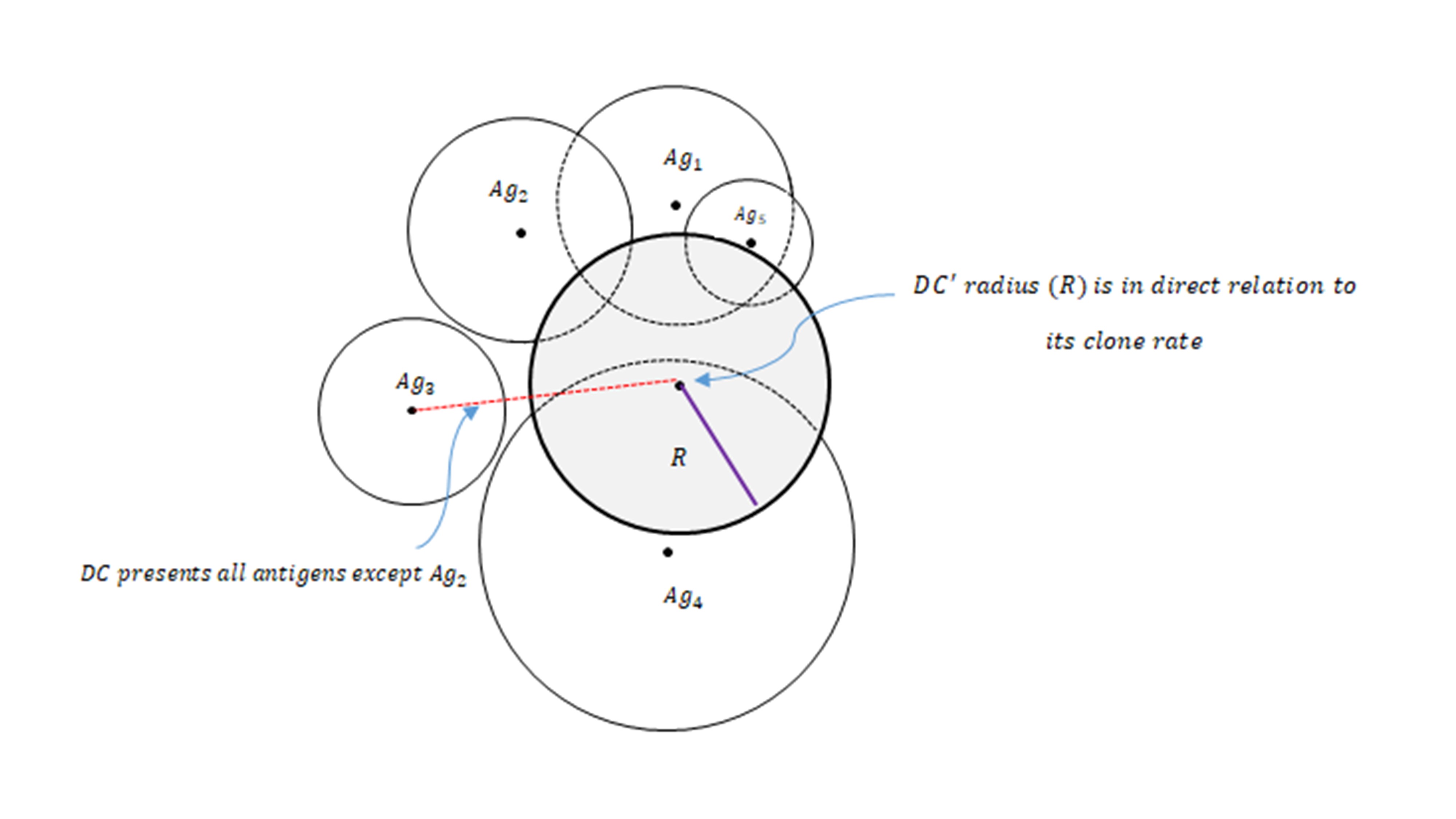}
\caption{\label{Fig.3} An illustration for non-randomly sampling antigens by DC on ${i}^{th}$  tissue, DC locates on $Ag_{6}$’s position and begins to check up coverage by sampling}
\end{figure}

In order to propose a semi-supervised classifier based on the DT concept, we improved the DCA original version proposed in Ref.\cite{Ref16}.   Although, it is not tailored to ID problem by default, it does not have some complexities of “process strategy” against the most of meta-heuristic algorithms like ACO, BHA, PSO, GA and etc. All of them solve ID problem by a centralized approach \cite{Ref59,Ref60,Ref61,Ref62,Ref63,Ref76,Ref77}; to search for finding the best optimum centroids among candidate solutions. Tests have accomplished in Sec.5 confirm this fact. The immune inspired DCA’s decentralized processing method in the binary classification is the reason why we adopted it as a defense shield and a good option for our purpose. To identify attacks at least with a tolerable false positive error is promising.  Calculating the DS has been done differently from two other signals in Ref.\cite{Ref16}.  It needs the incoming data be pre-ordered on class column while entails to a challenge. Thus, the danger signal processing is very important since it has a direct relevancy to pass false negative errors. When greater number of antigens are incorrectly recognized to normality, the possibility of FNR will be higher. In our mind, both of normal and abnormal samples of the train set (self) in addition of new labeled antigens which can be easily used to address this problem if we disregard the analysis of these innate features from traffic flow.
\\
\paragraph{\textbf{An idea }}   We used self-data in order to obviate challenge of being preordered. At first, minimum Euclidean distances between each antigen and all normal and abnormal selves are calculated separately. Then, antigen will be labeled to abnormal and added it to self-abnormal set if its minimum distance from the self-normal set is more than its minimum distance from the self-abnormal (Alg.2), and vice versa. Also, as stated in Sec\ref{2.1}, calculated CSM signal can be a good measure to check migration terms of DCs since it is computed through weighted sum equation (Eq.\ref{m1}).  Thus, the MT value must be predetermined in each run cycle of algorithm. But, it has usually a constant value in all phases of the DCA runs. So, some questions may raise as follows: 
\\
\paragraph{\textbf{Question }}  Could the MT be merely dependent on the subset of attributes mapped to the input signals and be dynamically provided meantime each running cycle of the DCA?  If so, will it impress classification performance? 
\\
\paragraph{\textbf{An idea }} Flexibility for determining the MT. In Alg.2 (Fig.5 at appendix), a constant MT value which is determined by user in range of [0, 1] for all run cycles, is requisite for migration of DCs. It will take place only if the CSM values exceed MT. As regards, the CSM calculation relies on input signals based on weight coefficients and also each dendritic cell has a different CSM value, it seems that the MT mostly takes effect from processed input signals which are variable cell by cell and cycle by cycle. Therefore, necessity of the MT‘s flexibility is required. So, Eq.\ref{mt}  is proposed to compute the MT value dynamically. Its proof for practicality is performed in experiment section. It is defined as below: 
\\
\begin{equation}\label{mt}
\footnotesize\ mt=[(F(P_{PAMP}) \times 2)+(F(P_{SS}))+(F(P_{DS}) \times 2)] 
\end{equation}
\\
Where, $P_{PAMP} , P_{SS} , P_{DS}$  are probabilities of three input signals for immature DCs in real range of [0, 1].  Also, either mean or median criteria can be taken for F function. Coefficients are respectively 2, 1 and 2 refer to WCM (Eq.2)
\\
\paragraph{\textbf{An idea }}  Procedure of sampling multiplied antigens vector \footnote{Vector are created from network traffic samples}. According to the Sec\ref{2.1} , DCs settle on positions of traffic samples when they want to inspect. Though, no strategy or idea has been adopted on the $Get_{Antigen()}$ function but randomly sampling entails the DCA to provide undesirable results in several cycles in each time. This leads to uncertainty. Whereas it must be accomplished by biological behaviors of DCs in tissues, we proposed a novel strategy which exactly acts based on its immunological functionalities of dendritic cells as below. Its proof for practicality is performed in the experiment section. It is defined as below, Calculating radiuses of coverage:
\\
\begin{eqnarray}\label{DCEq}
\begin{gathered}
\footnotesize\ 
Present_{radiusDC} = [\frac{Size_{of}Vector(j)}{max(Size_{of}Vector(j))}] \times F(dist(DC_{j},i))  \\
Size_{of}Vector(j) = [Rand(minCloneRange, maxCloneRange) \times \\
Size_{of}tissues_{j}] + Size_{of}tissues_{j}
\end{gathered}
\end{eqnarray}
\\  
Where, $Size_{of}tissues_{j}$  is number of antigens into ${j}^{th}$  tissue; $j \in [1,number\,of\,ce\\ll,tissues]$,  i  is ${i}^{th}$  antigen per  n  exist into  ${j}^{th}$  tissue, F can be either mean or median, $Size_{of}Vector(j)$ is also $Clone_{vector}$  rate as to ${j}^{th}$  tissue  includes n antigens to specify  $Antigen_{vector}$.  Clone indicator really means maximum length that vector can possess. Fig.\ref{Fig.3}  shows how each antigen is multiplied.  Eq.\ref{DCEq}  gives us a value that we can consider it as hyper shape radius which creates area about DC. So, sampling antigens by one DC depends on cell coverage.  The determination of radius of coverage depends on two things: (1) measuring Euclidian distance between inspector DC to other pervaded antigens, (2) The rate of multiplication for that inspected antigen. Actually, DC must sample antigens which have overlapping with them into its coverage. Fig.\ref{Fig.3}  clearly illustrates our strategy for sampling.
\\
\paragraph{\textbf{An example }}  
Assume that we have 10 tissues and the second one includes three types of antigens. Min and max ranges of antigen clone are also between 0.15 and 0.35. So, $Clone_{vector}$   is calculated as below: 
\begin{eqnarray}
\begin{gathered}
\footnotesize\ 
Size_{of}tissue(2) \leftarrow 3, \nonumber \\
Clone_{vector}(j=2) \leftarrow [Rand(0.15, 0.35) \times 3] + 3, \nonumber \\
\underrightarrow{yields}\ \ [\underbrace{Rand(0.15, 0.35)}_{0.25} \times 3] + 3 \rightarrow 3.75 \approx 4; \ \ \  3 \le Clone_{vector}(2) \le 6 \nonumber
\end{gathered}
\end{eqnarray}
Now, Distributing $Size_{of}Vector(i)$ into antigens: 
\begin{eqnarray}
\begin{gathered}
\footnotesize\ 
IF\, \left\{\begin{array}{lr} dist(Ag_{1}\,to\,Ag_{2}) \leftarrow 0.17 \\ dist(Ag_{1}\,to\,Ag_{3}) \leftarrow 0.36 \\ dist(Ag_{2}\,to\,Ag_{3}) \leftarrow 0.59  \end{array}\right. \Rightarrow \left\{\begin{array}{lr} Present_{radius}DC1 = [\frac{4}{6}] \times \overbrace{mean}^{F}(0.17, 0.36) \rightarrow 0.1767 \\ Present_{radius}DC2 = [\frac{4}{6}] \times mean(0.17, 0.59) \rightarrow 0.2534 \\ Present_{radius}DC3 = [\frac{4}{6}] \times mean(0.36, 0.59) \rightarrow 0.6334] \end{array}\right.  \nonumber \\
\underrightarrow{yields} \ \left\{\begin{array}{lr} \{Ag_{2}. Ag_{3}. Ag_{3}. Ag_{3}\}\,are\,sampled\,by\,DC_{1} \\ \{Ag_{1}. Ag_{3}. Ag_{3}. Ag_{3}\}\,are\,sampled\,by\,DC_{2} \\ \{Ag_{1}. Ag_{1}. Ag_{1}. Ag_{2}\} \,are\,sampled\,by\,DC_{3} \end{array} \right.\nonumber\\
\underrightarrow{yields} \ \left\{\begin{array}{lr} r_{Ag2} \leftarrow  \frac{(0.1767-0.17)}{[(0.1767-0.17)+(0.1767-0.36)]} \times 0.17 \rightarrow 0.006, \\ r_{Ag1} \leftarrow \frac{(0.2534-0.17)}{[(0.2534-0.17)+(0.2534-0.59)]} \times 0.17 \rightarrow 0.0337, \\ r_{Ag1} \leftarrow \frac{(0.6334-0.36)}{[(0.6334-0.36)+(0.6334-0.59)]} \times 0.36 \rightarrow 0.3107,  \end{array} \right. \nonumber\\
; \left\{\begin{array}{lr} r_{Ag3} \leftarrow \frac{(0.1767-0.36)}{[(0.1767-0.17)+(0.1767-0.36)]} \times 0.36 \rightarrow 0.347 \\ r_{Ag3} \leftarrow \frac{(0.2534-0.59)}{[(0.2534-0.17)+(0.2534-0.59)]} \times 0.59 \rightarrow 0.472 \\ r_{Ag2} \leftarrow \frac{(0.6334-0.59)}{[(0.6334-0.36)+(0.6334-0.59)]} \times 0.59 \rightarrow 0.081  \end{array} \right. \nonumber\\
\end{gathered}
\end{eqnarray} 
\\
According to Alg1. line of 4, antigen set is divided into cell tissues. Two nominal attributes of protocol and service of UNSW-NB15 dataset are utilized for this purpose. They have been highlighted on Tab.7 at appendix. \\
\begin{algorithm}[H]
\caption{The proposed pseudo code for data sampling }
  \algsetup{linenosize=\tiny}
  \scriptsize\begin{algorithmic}[1]
\STATE\textbf{Input:} S=$Antigen_{Set}$ . Mn=minAgCloneSz . Mx=maxAgCloneSz . Sg=Signal . Pr=$Probabilities_{of}InputSignalValues$ . MT=$Migration_{Threshold}$
\STATE\textbf{Output:} E=$Set_{of}MigratedCells$ . U=$Set_{of}UnmigratedCells$ . C=$Cell_{Context}$
{\color{blue}\STATE $/*$ Detection Phase $*/$}
\STATE $tissues$=$divide$ $S$ $to$ $cell$ $tissues$ $based$ $on$ $Protocol-Service$ $attributes$
\FOR {\textbf{each} tissues}
\STATE $Size_{of}Vector(i)=[RANDOM(Mn,Mx) \times Size_{of}tissues(i)]+Size_{of}tissues(i)$
\FOR {\textbf{each} selected DC into tissues(i)}
\STATE $-$ $Get$ $the$ $Antigen:$ $Distance$ $Antigens$ $to$ $DC$ $\gets$ $Determine$ $Euclidian$ $distance$ $DC$ $from$ $antigens$ $into$ $tissue(i).$
\STATE $-$ $Distributing$ $Size_{of}Vector(i)$ $into$ $antigen$ $is$ $toward$ $DC$ $coverage,$ $its$ $clone$ $rate$ $will$ $be$ $higher$ $and$ $vice$ $versa.$
\STATE $R \gets Calculate$ $DC$ $radiuses$ $utilizing$ $\textbf{Eq.6}$
{\color{blue}\STATE $/*$ Preprocessing antigens by DC $*/$}
\FOR {\textbf{each} antigens of tissues(i)}
{\color{blue}\STATE $/*$ Overlapping between DC $\&$ Ag $*/$}
\IF {$(R+r_Ag(i)) > Dist(DC from Ag(i))$} 
{\color{blue}\STATE $/*$ Signal Assessment $*/$}
\STATE $-$ $Get$ $the$ $Sg$ 
\STATE $-$ $Calculate$ $interim$ $output$ $signals$ $(\textbf{Eq.1})$
\STATE $-$ $Update$ $the$ $cumulative$ $output$ $signals$ $(\textbf{Eq.1})$
{\color{blue}\STATE $/*$ requisite for migration $*/$}
\IF {$cumulative$ $CSM$ $>$ $MT$}
\STATE $-$ $Remove$ $the$ $DC$ $population$
\STATE $-$ $Assign$ $the$ $C$ $to$ $DC$
{\color{blue}\STATE $/*$ Context assessment phase $*/$}
\IF {$cumulative Semi \leq cumulative Mat$}
\IF{$Cell_{Context} \geq 0.50$}
\STATE $-$ $Add$ $DC$ $to$ $E.Mat_{cells}$
\ENDIF
\ELSE
\IF {$Cell_{Context} < 0.50$}
\STATE $-$ $Add$ $DC$ $to$ $E.Semi_{cells}$
\ENDIF
\ENDIF
\STATE $-$ $Termination$ $the$ $DC$ $and$ $Add$ $a$ $naive$ $DC$ $to$ $the$ $population$
\ELSE 
\STATE $-$ $Add$ $DC$ $to$ $U$ $to$ $sample$ $again$
\ENDIF
\STATE $-$ $DC$ $back$ $to$ $population$
\ENDIF
\ENDFOR
\ENDFOR
\ENDFOR
\end{algorithmic}
\end{algorithm}

\begin{algorithm}[H]
\caption{The proposed pseudocode to make antigen set preordered}
  \algsetup{linenosize=\tiny}
  \scriptsize
\begin{algorithmic}[1]
\STATE\textbf{INPUT:} S=$Antigen_{Set}$ . SN=$Self_{Normal}$ . SA=$Self_{Abnormal}$
\STATE\textbf{OUTPUT:} $O=PreOrdered_{Antigen_{Set}}$
\FORALL {$Ag_{i} \in S$}
\STATE $Candidate_{distN}=min_{Euclidian\,distance}(Ag_{i}\,from\,SN)$
\STATE $Candidate_{distA}=min_{Euclidian\,distance}(Ag_{i}\,from\,SA)$
\IF {$Candidate_{distN} > Candidate_{distA}$}
\STATE $Ag_{i}LABLE \leftarrow Abnormal$
\STATE $Add\,Ag_{i}\,to\,SA$
\ELSE
\STATE $Ag_{i}LABLE \leftarrow Normal$
\STATE $Add\,Ag_{i}\,to\,SN$
\ENDIF
\ENDFOR
\STATE $O=SORT\,\,'S'\,\,based\,on\,Class labels$
\STATE $RETURN\,O$
\end{algorithmic}
\end{algorithm}
\ \\
According to Alg.2, unlabeled antigens with labeled ones are used to make antigen set as preordered. Therefore, two candidate distance values are calculated based on Euclidian distance between $Ag_{i}$ from self-sets. We need this process to generate danger signal by Information Gain (IG) method (Line 5 from Alg.3). In tests, we adopted IG to compute input signals. Two SS and PAMP signals are obtained in the same way, but the DS has a difference strategy. Ref.\cite{Ref16} has been precisely described how to compute these three input signal values through IG. We subsequently proposed our improved DCA. \\
\\
\begin{algorithm}[H]
\caption{Pseudocode of the proposed DCA}
  \algsetup{linenosize=\tiny}
  \scriptsize
\begin{algorithmic}[1]
\STATE\textbf{Input:} S=$Antigen_{Set}$ . T=$Train_{Set}$ . $Mn=minAgCloneSz$ . $Mx=maxAgCloneSz$ . $Weight_{Matrix}$  . $Sg=Signal$ . Pr=$Probabilities_{of}InputSignalValues$ . MT=$Migration_{Threshold}$
\STATE\textbf{Output:} E=$Set_{of}MigratedCells$ . MCAV
{\color{blue}\STATE $/*$ Preprocessing $*/$}
\STATE $-$ $Sort$ $S$ $to$ $make$ $it$ $preordered$ \textbf{(Alg.2)} 
\STATE $-$ $Select$ $best$ $method$ $to$ $compute$ $input$ $signals.$ $i.e.$ $the$ $information$ $Gain$ $(IG)$
\STATE $-$ $Compute$ $MT$ $adopting$ $Pr$ $and$ $Weight$ $Coefitients$ $(\textbf{Eq.2}$ $and$ $\textbf{Eq.5})$ 
{\color{blue}\STATE $/*$ Detection Phase $*/$}
\WHILE {U is not empty}
\STATE $[E, U, C] \leftarrow Sampling\,Antigens(S, Mn, Mx, Sg, Pr, MT)\,\textbf{Alg.1}$
{\color{blue}\STATE $/*$ Classification phase $*/$}
\FORALL {$E.Semi_{cells}$}
\STATE $Semi_{DCs}(Ag_{i}) \leftarrow \frac{Semi_{cells}(Ag_{i})-min(Semi_{cells})}{max(Semi_{cells})-min(Semi_{cells})}$
\ENDFOR
\FORALL {$E.Mat_{cells}$}
\STATE $Mat_{DCs}(Ag_{i}) \leftarrow \frac{Mat_{cells}(Ag_{i})-min(Mat_{cells})}{max(Mat_{cells})-min(Mat_{cells})}$
\ENDFOR
\FORALL {$member of E$}
\IF {$C(i) \geq 0.50$}
\STATE $MCAV(i) \leftarrow mean(\frac{Mat_{DCs}(i)}{Mat_{DCs}(i)+Semi_{DCs}(i)},\frac{E(i).Present}{E(i).Clone})$
\ELSE
\STATE $MCAV(i) \leftarrow mean(\frac{Semi_{DCs}(i)}{Mat_{DCs}(i)+Semi_{DCs}(i)},\frac{E(i).Clone-E(i).Present}{E(i).Clone})$
\ENDIF
\ENDFOR
\ENDWHILE

\end{algorithmic}
\end{algorithm}

Regarding Alg.3, there are seven input and two output parameters. In each iteration cycle of the algorithm, we have two calls related with algorithms 1 and 2 on lines 4 and 9 that algorithm 1 (sampling antigens function) takes a number of input parameters. Among these parameters, the MT and the input signal probabilities are calculated as variables in each run cycle and have some dynamic value. We also made changes to the MCAV calculation that are observable. In the case of parameter setting, MT, minAgCloneSz and maxAgCloneSz parameters are single values where the MT value is dynamically changed according to the size of the CSM signal in each run cycle of the whole algorithm. A significant point is clone size of antigens which we can determine a range for them. If this range is [a,b] that a is minAgCloneSz and b is maxAgCloneSz, this means that a hypothetical antigen’s clone rate can be random variable between a and b. In addition, the parameter of Pr determines the probability of input signals in the range [0,1] as a percentage in Algorithm 1. At the output of the Sampling antigens function, the three parameters E, U, C respectively specify the values of the migrated, non-migrated cells and the context values for the cells. Algorithm 3 continues until the size of the set U is zero and all cells migrate, so the number of execution cycles depends on the size of this parameter. Fig.4 illustrates the general procedure of the proposed method. 
\\
\begin{center}
\footnotesize\
\includegraphics[scale=0.47]{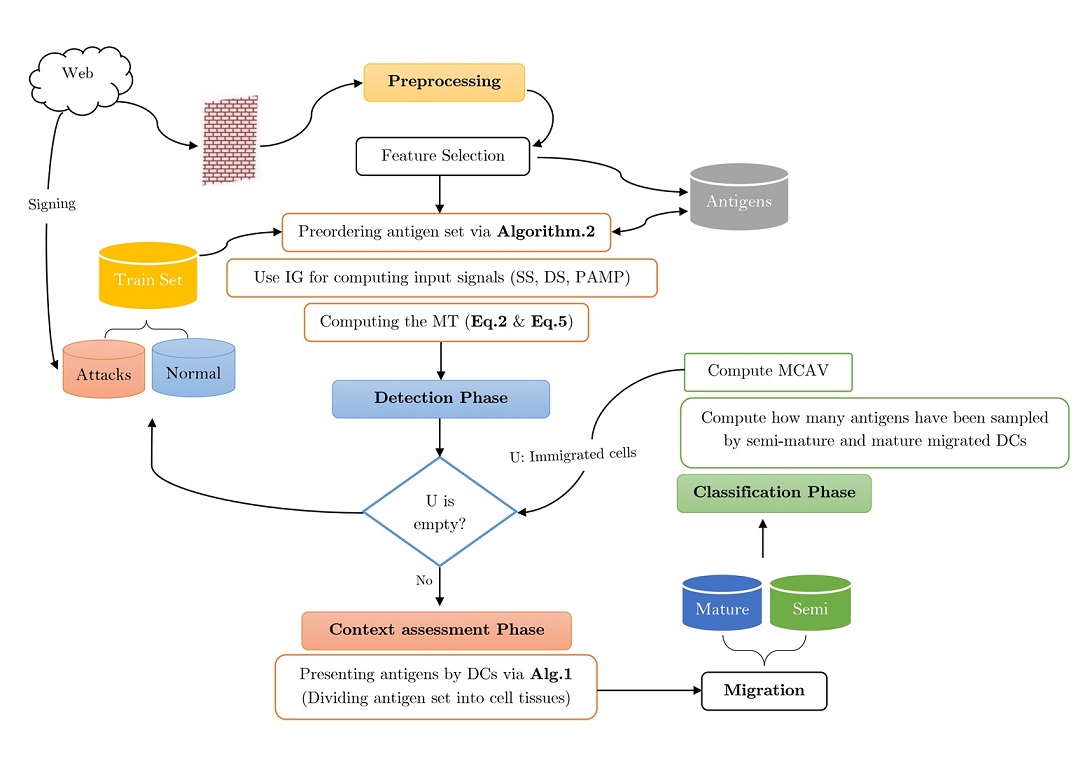} \\ 
\textbf{Fig.4}  General procedure of the proposed method \\
\end{center}

\section{Experimental and Results} 
In order to evaluate our proposed defensive shield, in this section, the required simulation was accomplished with traffic samples uniformly random derived from a reliable network intrusion dataset, the UNSW-NB15 dataset provided by University of New South Wels \cite{Ref31}.  The involved attacks of the UNSW-NB15 data set were categorized into nine types as following: Fuzzers, Analysis, Backdoor, DoS, Exploit, Generic, Reconnaissance, Shellcode, and Worm. Table.7 at appendix shows its attributes in details. Also, the involved features of this dataset are classified into six groups as follows: flow, basic, time, additional generated and labelled features. And in total, it has 49 attributes. It is one of the newest IDS datasets which has been recommended to use in ID problems instead of the former datasets like KDDCup99. Our proposed technique is simulated with a classic computer Intel® Core ™ i5-3230M CPU @ 2.60Hz, 4.00 GB Ram equipped with the MATAB R2018b environment. 
\subsection{\label{4.1} Evaluation criteria}
Following performance measures were utilized for evaluating the proposed method: The Detection rate (DR), False Positive Rate (FPR), False Negative Rate (FNR), and Accuracy Rate (ACC) and complements of them in addition to Correlation Coefficient (cc) and CID which relatively appears unfamiliar in this case \cite{Ref36}. 
\begin{eqnarray}
\footnotesize
\begin{gathered}
DR=\frac{No\,of\,attacks\,that\,are\,correctly\,classified\,as\,attack}{Total\,No.\,of\,attacks\,in\,the\,test\,dataset} \times 100 \\
FPR=\frac{No.\,of\,normal\,that\,are\,uncorrectly\,classified\,as\,attack}{Total\,No.\,of\,normal\,in\,the\,test\,dataset} \times 100\\
FNR=\frac{No.\,of\,anomaly\,that\,are\,uncorrectly\,classified\,as\,normal}{Total\,No.\,of\,anomaly\,in\,the\,test\,dataset} \times 100\\
ACC=\frac{No.\,of\,correctly\,classified\,as\,samples}{Total\,No.\,of\,samples\,in\,the\,test\,dataset} \times 100\\
cc=\frac{(TP\times TN)-(FP \times FN)}{\sqrt{(TP+FN)\times (TP+FP)\times (TN+FP)\times(TN+FN)}} \\
\end{gathered}
\end{eqnarray}
\\Where,  $cc \in [-1, 1]$.  The cc value closer to -1 indicates that the predictive rate for the classification performance is inconsistent with reality. According to the information theory-based equations given on Ref.\cite{Ref36},  Simplification of CID depends on three NPV \footnote{Negative predictive value, $P(\lnot I|\lnot A)$  is the chance that there is no intrusion, $\lnot I$,  when an IDS does not output an alarm, $\lnot A$}  , PPV \footnote{Positive predictive value, $P(I|A)$  is the chance that an intrusion, $I$  is present when an IDS output an alarm $A$ } and Base (B \footnote{Base rate,  $P(I)$  is the probability that there is an intrusion in the observed audit data,  $B$  rate is consider $10^{-5}$ by default}) rates. \\
\begin{eqnarray}
\footnotesize
\begin{gathered}
PPV \leftarrow P(I|A)=\frac{P(I, A)}{P(A)}+\frac{P(I)P(A|I)}{P(I)P(A|I)+P(\lnot I)P(A|\lnot I)} \\
NPV \leftarrow P(\lnot I|\lnot A)=\frac{(1-B)(1-\alpha)}{(1-B)(1-\alpha)+B\beta}\\
\end{gathered}
\end{eqnarray}
Where, I and A are respectively intrusion and alarm,  $\alpha , \beta$  are FP and FN. By replacing quantities to these relations, we simplified them as bellows to be utilizable:
\begin{center}
\begin{eqnarray}
\footnotesize
\begin{gathered}
PPV=\frac{B. TP}{B. TP+\bar{B}. FP},\ \ \ NPV=\frac{(1-B).(1-FP)}{(1-B)(1-FP)+B. FN}\\
\\ C_{ID}=-B(1-\beta)log(PPV)-B(1-\beta)log(1-NPV)-\\(1-B)(1-\alpha)log(NPV)-(1-B)\alpha log(1-PPV) \\
\end{gathered}
\end{eqnarray}
\end{center}\
\\But, CID was still able to be simplified more:
\begin{center}
\begin{eqnarray}\label{Eq13}
\footnotesize
\begin{gathered}
C_{ID}=-B(1-FN)log(\frac{B. TP}{B. TP+\bar{B}. FP})-\\ B(1-FN)log(1-\frac{(1-B).(1-FP)}{(1-B)(1-FP)+B. FN})-\\(1-B)(1-FP)log(\frac{(1-B).(1-FP)}{(1-B).(1-FP)+B.FN})-\\(1-B)FPlog(1-\frac{B. TP}{B. TP+\bar{B}. FP}) 
\end{gathered}
\end{eqnarray}
\end{center}
\paragraph{\textbf{Preparation of dataset}}
We extracted subsets with sizes of 2110 and 20000 network traffic records from UNSW-NB15 respectively for test and train phases in our experiments. In order that traffic samples would be selected uniformly random, we used the MATLAB environment. Moreover, the min-max normalization technique was applied due to the nominal features which can be turned into the numerical ones and normalized in range of [0,1]. Besides, we embedded a number of unseen attacks (4.075 percentage means 86 records) into test subset so as to try unseen detection rates in our proposed system.\\
\subsection{\label{4.2} Simulating the behavior of DCs}
This part mainly studies on improvement of the standard DCA as a new semi supervised method in ID problem adopting the proposed ideas. In previous section, it is mentioned that meta-heuristic techniques aren’t appropriate to be used in binary classification of anomalous traffic due to their detection strategy which is centroid based. As real traffic samples often make entangled clusters with a high density in problem n-dimensional space causes the binary classification is impossible. So, incoming data can be divided into multiple normal and abnormal clusters for doing non-binary classification. In most cases, non-linear methods cannot ever do this task effectively unlike claim posed in Ref.\cite{Ref34}.  According to performed tests in this section, this claim will not be true in case of “network anomaly detection”. Hence, utilizing the AIS distributed methods can be very beneficial. In continue, yields have gotten by comparisons indicate this fact.
Following test stages show improvement procedure of our proposed algorithm to make this defensive shield against intrusion.
\\
\begin{center}
\footnotesize\textbf{Table.1}      Allocate input parameters of the standard DCA (Fig.8 at appendix), a used-defined MT and random sampling in 8 times of run cycles 
\footnotesize\renewcommand{\arraystretch}{0.85}
\begin{longtable}{|c|c|ccccc|}
\hline
{\textbf{No.}}&{\textbf{Selected features}}&{\textbf{mt}}&{\textbf{at}}&{\textbf{runtime (s)}}&\rotatebox{90}{\textbf{minAgClone}}&\rotatebox{90}{\textbf{maxAgClone}}\\\hline
1&All&50&50&26.4219&5&10\\
2&All&40&40&26.4219&5&15\\
3&All&40&50&17.143&5&25\\\hline
4&Best Features: 8.9.30.33.41.42   by ref\cite{Ref31}&50&50&15.8608&5&10\\
5&''&35&25&15.3479&10&40\\\hline 
6&Middle ranked features:&50&50&26.7069&5&10\\
&4-6.10-11.15.18-26.31-32.40  by ref\cite{Ref31}&&&&&\\
7&''&35&35&16.1472&10&40\\\hline 
8&Achieved by CFA method (ref\cite{Ref37})&40&40&19.0889&3&10\\
&1-6.8-11.14.18-24.30.32.33.41.42&&&&&\\\hline
\end{longtable}

\footnotesize\textbf{Table.2}      Perf. of the standard DCA (Fig.8 at appendix) in 8 times of run cycles\\
\begin{longtable}{|c|ccccccc|}
\hline
&&&&&misClassification&Correct&Correct\\
Run.&DR&FPR&FNR&ACC&rate&prediction&prediction\\
&&&&&&normal&attack\\\hline
1&51.4&46.5&48.6&52.0&48.0&33.4&70.8\\
2&48.2&34.8&51.8&53.5&46.5&36.4&75.2\\
3&46.3&33.9&53.7&52.5&47.5&35.9&75.0\\
4&51.0&34.1&49.0&55.6&44.4&38.0&76.7\\
5&39.4&24.5&60.6&50.7&49.3&36.2&77.9\\
6&54.3&42.3&45.7&55.4&44.6&36.5&73.9\\
7&45.9&27.9&54.1&54.1&45.9&37.7&78.3\\
\textbf{8}&\textbf{62.6}&\textbf{44.5}&\textbf{37.4}&\textbf{60.4}&\textbf{39.6}&\textbf{40.3}&\textbf{75.5}\\
\hline
\end{longtable}
\end{center}
According to Alg.3 (line of 6), we utilized Eq.5 for computing variable MT by 20 running cycles. So, the best configuration of input parameters of Alg.3 is obtained as follows:” Mean” for F in Eq.5 to compute MT, features of ”1-6, 8-11, 14, 18-24, 30, 32-33, 41-42” for selected optimum features by utilizing CFA, 50 for 'at' parameter, [5, 15] for range of antigen clone sizes. Results of running Alg.3 through this configuration is within Tab.8. In general, we tested four approaches of the proposed DCA have been specified by four colors. (Fig.7) \\ \ \\
\begin{center}
\footnotesize\textbf{TABLE 3.}      Yields for the proposed DC algorithm
\begin{longtable}{p{0.7cm}|p{0.5cm}p{0.5cm}p{0.5cm}p{0.5cm}p{0.5cm}p{0.5cm}p{0.5cm}p{0.5cm}p{0.5cm}p{0.5cm}p{0.5cm}p{0.5cm}p{0.5cm}p{0.5cm}}
\hline
\rotatebox[origin=c]{270}{\textbf{Meth. (as Color)}}&\rotatebox[origin=c]{270}{\textbf{DR}}&\rotatebox[origin=c]{270}{\textbf{FPR}}&\rotatebox[origin=c]{270}{\textbf{FNR}}&\rotatebox[origin=c]{270}{\textbf{TP}}&\rotatebox[origin=c]{270}{\textbf{FP}}&\rotatebox[origin=c]{270}{\textbf{FN}}&\rotatebox[origin=c]{270}{\textbf{Accuracy}}&\rotatebox[origin=c]{270}{\textbf{Error}}&\rotatebox[origin=c]{270}{\textbf{F-measure}}&\rotatebox[origin=c]{270}{\textbf{Precision rate}}&\rotatebox[origin=c]{270}{\textbf{Attack corrected prediction rate}}&\rotatebox[origin=c]{270}{\textbf{Normal corrected preiction rate}}&\rotatebox[origin=c]{270}{\textbf{Attack in corrected prediction rate }}&\rotatebox[origin=c]{270}{\textbf{Normal in corrected prediction rate}}\\\hline
Green&95.0&56.4&5.0&65.3&17.6&3.4&79.0&21.0&85.74&78.77&78.7&80.0&21.3&20.0\\
Blue&\textbf{86.2}&\textbf{20.3}&\textbf{13.8}&\textbf{59.2}&\textbf{6.4}&\textbf{9.6}&\textbf{84.2}&\textbf{15.8}&\textbf{88.17}&\textbf{90.24}&\textbf{90.3}&\textbf{72.5}&\textbf{9.7}&\textbf{27.6}\\
Red&65.7&37.4&34.3&45.14&25.70&10.72&64.7&35.3&64.70&63.72&45.4&79.4&54.6&20.6\\
Violet&94.5&57.0&5.5&64.9&17.8&3.8&78.4&21.6&85.74&78.48&78.5&78.0&21.5&22.0\\
Light Blue&\textbf{95.5}&\textbf{55.2}&\textbf{4.5}&\textbf{65.6}&\textbf{17.3}&\textbf{3.1}&\textbf{79.7}&\textbf{20.3}&\textbf{84.54}&\textbf{79.13}&\textbf{79.2}&\textbf{82.0}&\textbf{20.8}&\textbf{18.0}\\
Orange&62.6&44.5&37.4&43.0&13.9&25.7&60.4&39.6&68.47&75.57&75.5&40.3&24.5&59.7\\
\hline
\end{longtable}
\footnotesize\textbf{TABLE 4.}      Yields for the proposed DCA (Alg.3) by enhancing size of the AG set ten times 
\footnotesize\begin{longtable}{p{0.8cm}p{0.7cm}|p{1.2cm}p{0.6cm}p{0.6cm}p{0.6cm}p{1.2cm}p{0.6cm}p{0.6cm}p{1.3cm}p{1.3cm}}
\hline
&\textbf{Test Size}&\textbf{Runtime (S)}&\textbf{Dr}&\textbf{FP}&\textbf{FN}&\textbf{No. of false negative records}&\textbf{ACC}&\textbf{ERR}&\textbf{Total rate of unseen attacks (TP)}&\textbf{Non-detected unseen attacks (FN)}\\\hline
Light blue&20000&41.007&94.7&16.6&5.3&77&79.8&20.2&98.76&37 records\\
Curve&2110&28.368&94.5&57.0&5.5&86&78.4&21.6&79.07&18 records\\
\hline
\end{longtable}
\end{center}
At first, we simulated the standard DCA (Fig.5), and then tested it eight times under UNSW-NB15 dataset with 43 attributes on MATLAB environment. Details can be found on Tab.6 at appendix. In running cycles, the effectiveness of dimensional reduction on final performance was also analyzed. Although, it is tried to modify some of weight coefficients in WCM (Eq.\ref{mtrx}), it didn’t comprise good yields.\\
As it can be considered, obtained results (Tab.2) aren’t acceptable. Setting best input values are as follows: 40 for user-defined $mt$, 50 for 'at', and 5 and 10 respectively for min and max clone rates for antigen vector, as well as all of attributes are selected.
\begin{center}
\footnotesize\
\includegraphics[scale=0.4]{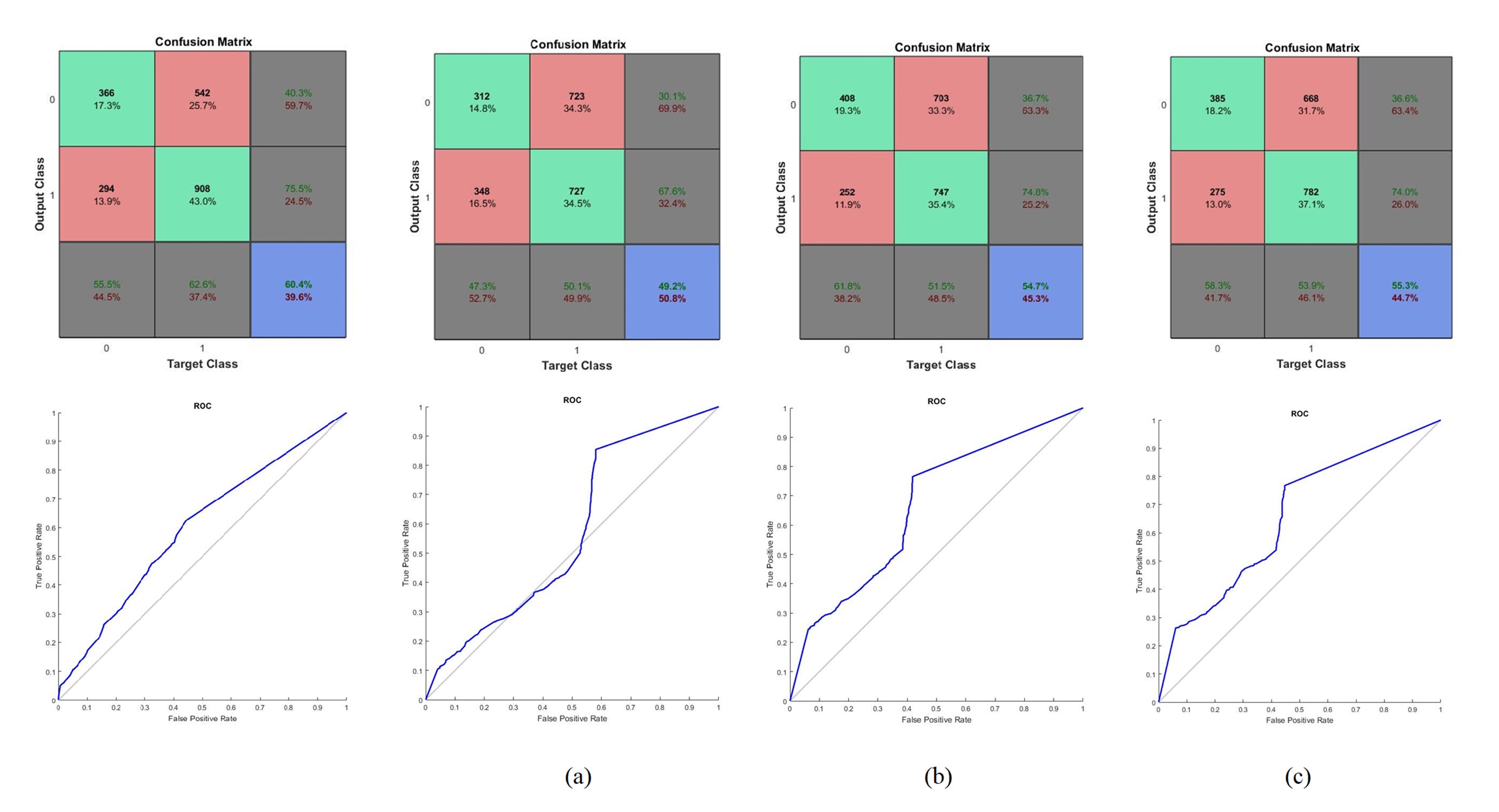} \\
\textbf{Fig.5}  Yields from three tests with applying  Eq.\ref{DCEq}  as radius strategy for sampling in non-randomly way. Function in a) is max b) is mean c) is median\\
\end{center}    
By analyzing tests, we employed “median” which can be a becoming standard for measuring coverage radius of dendritic cells. Details are available at Tab.6 at appendix. So, betterment in sampling is observable from (a) to (c) in Fig.5. We enforced another test, at this time by utilizing the CFA \cite{Ref37}  with ANN as learner classifier which is a powerful feature selection method inspired by cuttlefish mechanism presented by Sabry Eesa and et al. (Fig.6)\\
Here, it is necessary to provide an explanation of the values of the input parameters in Tab.6 and its effect on the output performance of the proposed system. Whatever the radius of presenting of DCs (rows 1 and 2 in Tab.6) be higher, the false positive error will be increased. As an instance, in row 2 of this table, the two maximum and mean criteria are respectively selected for the two parameters of radius of presenting of DCs and dynamic migration threshold, which its result is increase in detection rate and a significant decrease in false negative rate and an increase in accuracy and positive rates. Although the increase in false negative is highly undesirable, the classifier is successful when both of its false positive and false negative rates is low. Therefore, care must be taken in selecting the criterion type for these two sensitive input parameters. 
\begin{center}
\includegraphics[scale=0.22]{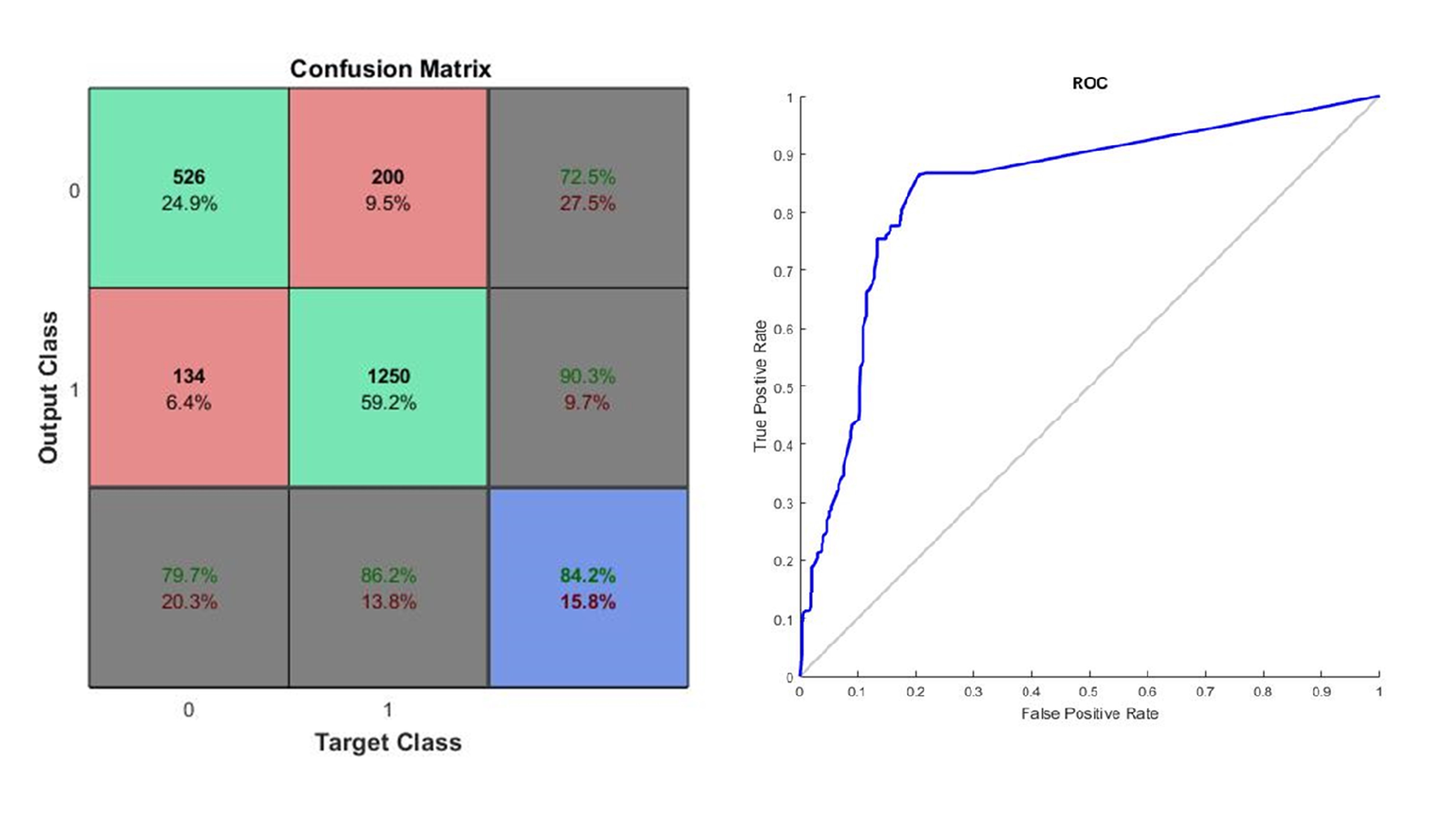} \\
\footnotesize\
\textbf{Fig.6} Curves belong to experiment of the proposed DCA (Alg.3) by utilizing Eq.\ref{DCEq} (F was median) and values from Tab.3\\
\end{center}
Notice that the “mean” was selected for F parameter in Eq.\ref{mt}. DCA Classification phase is finally performed by checking status of all migrated DCs. This means that it is checked for each AG and how many mature/semi-mature migrated cells have sampled it during DCA processes in detection phase. If number of mature cells be greater than semi-mature ones, it is likely that AG possesses anomalous pattern (attack). This probability will be higher when the difference between the number of mature and semi-mature cells is a lot. 
\\
Hitherto, we showed the effectiveness of our proposed ideas in Sec.4.  Fig.7 comprehensively exposes efficacy rate of proposed ideas in improvements of the original DC mechanism while sampling antigens are accomplished in both random and non-random ways (Eq.\ref{DCEq})  as compared to centroid based approaches like the Black hole clustering algorithm. It is proposed by Hatamlou and is actually inertia version of PSO which is widely used in classification problems. Curves with and without using the dimensional reduction method (CFA) also provide a better detection rate in condition of any prior training deficiency while this rate for BHA was 65.7. Whereas it is shown the DR of blue curve (86.2) is clearly a few lower than other approaches which Eq.\ref{DCEq}  has been used for them.  Chromatic blue curve belongs to the condition that non-random sampling was adopted. In like manner, the violet, green and light blue belong to simulated Alg.3. As shown from following Chart, it must be worked on distributed sampling proposed on Eq.\ref{DCEq}, especially blue curve, better than before.\\
So as to analyze unseen data detection rates (UDR), we enhanced the size of test subset from 2110 to 20000 records (tenfold). At gained set, the number of embedded unseen attacks were 3000 records (15 percentage). In test of light blue curve, UDR was 79.07 means just 18 from total 87 unseen anomalies which have been detected under previous test set that was not so ideal. According to Tab.4, 37 samples were detected as unseen anomalies from total 77 samples which wrongly were classified as normal profiles (FNR). Ordinarily, error rate due to detected unseen anomalies (UFNR) will be raised meantime increasing time and cost complexities while test size was increased. So, we showed raising UDR from 79.07 to 98.76 in Tab.4.\\
\begin{center}
\footnotesize\
\includegraphics[scale=0.45]{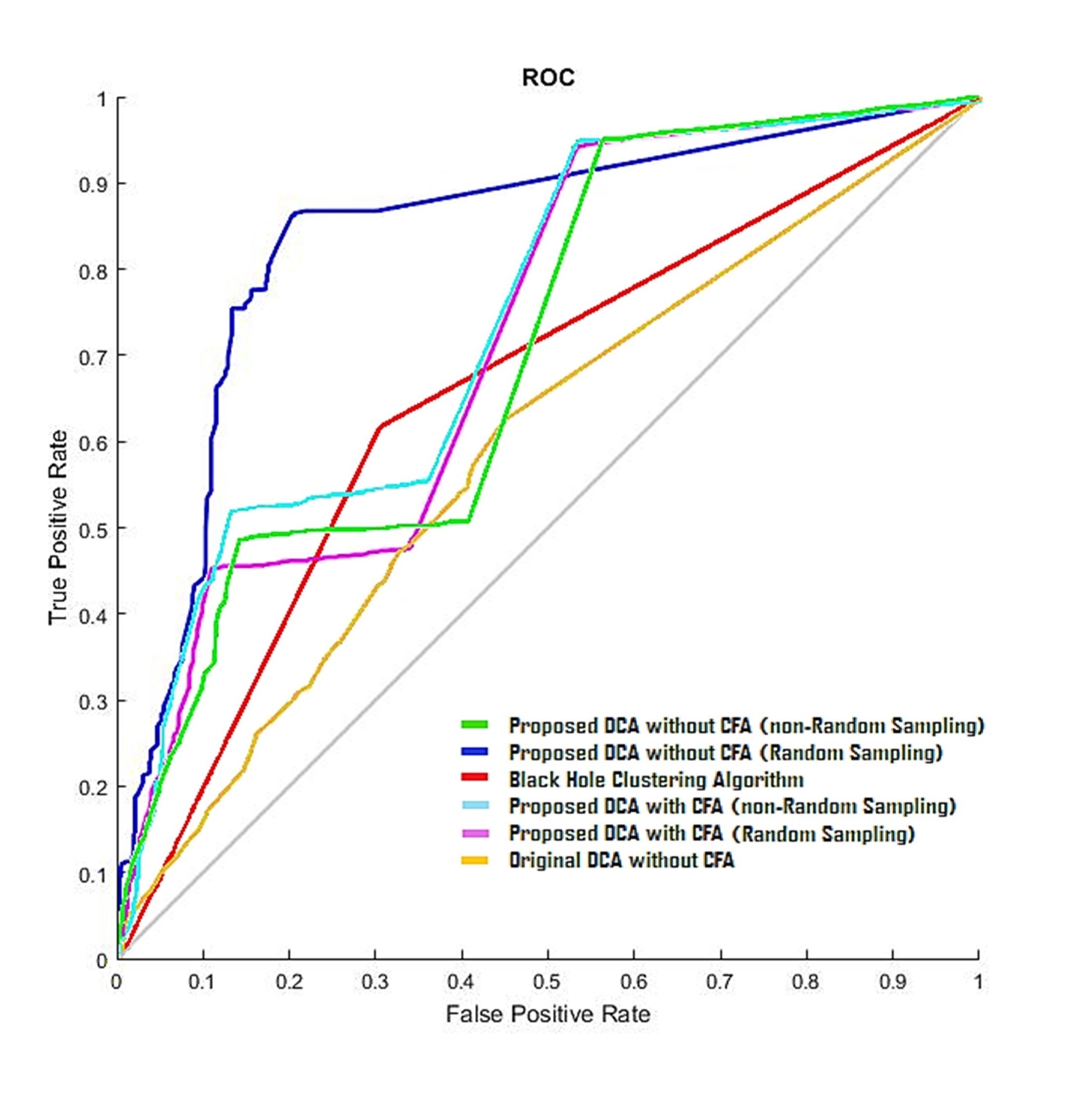} \\ 
\textbf{Fig.7}  Comparative results obtained from final experiment of the improved DCA (Alg.3)\\
\end{center}
\paragraph{\textbf{Analysis by CID criterion}}  
There is a criterion called “Capability of Intrusion Detection” that firstly was proposed in Ref.\cite{Ref36} and strongly recommended to be used in evaluations. It is used in cases that ROC curves cross each other like Fig.7, in such cases, both TP and FP are raised or reduced together so it is impossible to analyze which one of curves has better performance. It is defined in Eq.\ref{Eq13}.  Though it is based on information theory, we simplified it as follows so as to be more useable:\\
The probability of intrusion presence on test set, Base rate  $B=10^{-5}$  by default. Depending on Tab.5, the blue and the light blue with respectively 3.6443 and 10.4493 scores, are most capable among other methods. The use of CFA as the feature selection phase shows its influence on the superiority of the Light blue approach to green. In approaches with blue and violet colors, there is no method for sampling and it is random by default. The impact of the use of CFA on both green and light blue approaches has been positive in terms of CID rates. As contest between methods with violet, light blue and green colors and the influence of CFA usage in light blue curve compared with green ones is observable.\\

\begin{center}
\footnotesize\textbf{Table.5}  Competitive analysis of the DCA approaches \\
\footnotesize\begin{tabular}{c|c}
Method (by Color)&Capability of Intrusion Detection\\
\hline
Green&3.5753\\
\textbf{Blue}&\textbf{10.4493}\\
Red&2.3315\\
Violet&3.5311\\
\textbf{Light Blue}&\textbf{3.6443}\\
Orange&1.1769\\\hline
\end{tabular}
\\
\end{center}
 
\section{Conclusion and future directions}
In this research, we deeply studied improving the immune based DCA mechanism in anomaly detection problem. Its immunological background helped us much to inspiration. At this proposed semi-supervised classifier, DCs present antigens siting on their coverages non-randomly.  Mobility of migration threshold, non-random antigen sampling and necessity for antigen set to be  preordered are among problems that were addressed well. Concerning the results of the experiments, although the detection rate of the proposed approach using the non-random sampling idea was high, its intrusion detection capability is weaker than the default case of random sampling. However, detection rate of unseen attacks in the proposed method was also promising. But technically, achieving higher CID with higher DR is feasible, and this depends more on the application of the sampling criteria type (mean, median, etc.) as well as the correct evaluation of the input signals in the signal assessment phase. It shows that the proposed system still needs to be worked on and can complementary be a future plan for research.
\\

\section*{Acknowledgment }
\footnotesize This work derived from approved M.Sc. Thesis, is supported by Malek-Ashtar University of Technology (MUT), IRAN.  \\

\section*{Compliance with Ethical Standards}
\footnotesize\textbf{Conflict of interest:} On behalf of all authors, the corresponding author states that there is no conflict of interest. \\
\footnotesize\textbf{Funding:} This study is not funded by a specific project grant.\\
\footnotesize\textbf{Ethical approval:} This article does not contain any studies with human participants or animals performed by any of the authors.\\
\footnotesize\textbf{Informed consent:}  Informed consent was obtained from all individual participants included in the study.

\newpage
\section*{Appendix}
\begin{center}
\includegraphics[scale=1]{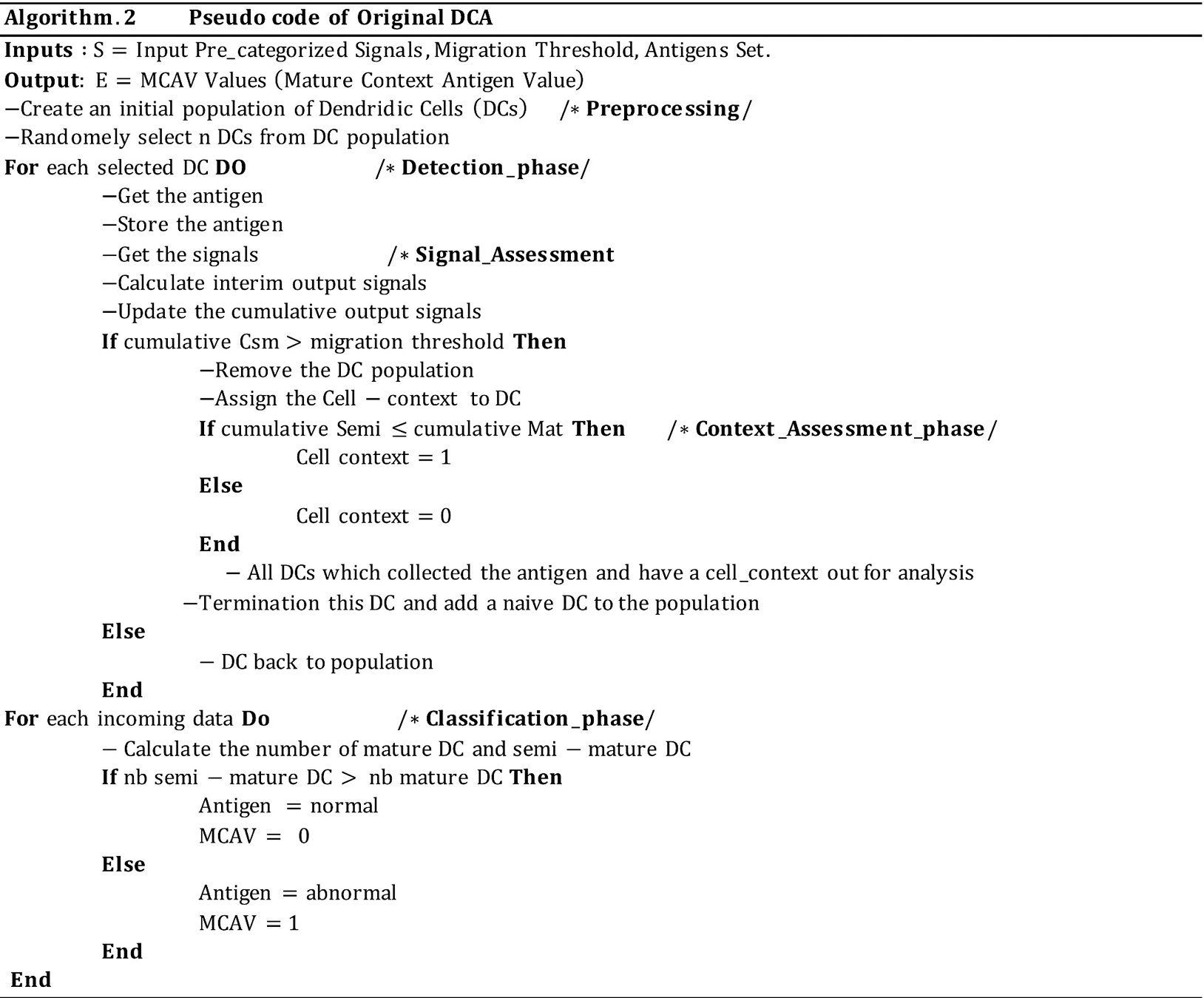}
\footnotesize\
\textbf{Fig.8} Pseudo code of the original version of the DCA, taken by Ref.\cite{Ref16} \\ \ \\
\footnotesize\textbf{Table.6}   Best configurations for eight approaches of the proposed DCA in the first antigen set (average 20 times of tests) \\ 
\footnotesize\renewcommand{\arraystretch}{1}
\begin{tabular}{|c|cccc|cccccccc|}
\hline
&\multicolumn{4}{|c|}{\textbf{Proposed DCA method}}&&&&&&&&\\\cline{2-5}
\multirow{2}{*}{\rotatebox[origin=c]{90}{\textbf{Test No.}}}&\rotatebox[origin=c]{90}{\textbf{minClone for presentedAg}}&\rotatebox[origin=c]{90}{\textbf{maxClone for presentedAg}}&\rotatebox[origin=c]{90}{\textbf{Radius of presenting DCs}}&\rotatebox[origin=c]{90}{\textbf{Migration Threshold (MT)}}&\multirow{2}{*}{\rotatebox[origin=c]{90}{\textbf{Iteration}}}&\multirow{2}{*}{\rotatebox[origin=c]{90}{\textbf{runtime (s)}}}&\multirow{2}{*}{\rotatebox[origin=c]{90}{\textbf{Recall}}}&\multirow{2}{*}{\rotatebox[origin=c]{90}{\textbf{FPR}}}&\multirow{2}{*}{\rotatebox[origin=c]{90}{\textbf{TNR}}}&\multirow{2}{*}{\rotatebox[origin=c]{90}{\textbf{FNR}}}&\multirow{2}{*}{\rotatebox[origin=c]{90}{\textbf{ACC}}}&\multirow{2}{*}{\rotatebox[origin=c]{90}{\textbf{ERR}}}\\\hline 
1&5&10&mean&median&21&0.178856&90.1&51.2&48.8&9.9&77.2&22.8\\
2&3&10&mean&median&21&0.176326&92.8&52.7&47.3&7.2&78.6&21.4\\
3&3&15&mean&median&21&0.223302&91.7&52.9&47.1&8.3&77.7&22.3\\
4&5&15&max&median&21&0.186085&93.6&57.6&42.4&6.4&77.6&22.4\\
5&5&15&max&mean&21&0.225727&97&63.2&36.8&3&78.2&21.8\\
6&5&15&median&mean&21&0.20068&95.5&62.9&37.1&4.5&77.3&22.7\\
7&5&15&median&median&21&0.181126&90.8&52.4&47.6&9.2&77.3&22.7\\
8&5&15&mean&mean&21&0.202256&95.1&55.8&44.2&4.9&79.2&20.8\\\hline
\multicolumn{13}{|c|}{Best proposed approach for setting DCA' input parameter}\\\hline
\textbf{9}&\textbf{5}&\textbf{15}&\textbf{AVG}&\textbf{median}&\textbf{21}&\textbf{0.255311}&\textbf{95.9}&\textbf{59.2}&\textbf{40.8}&\textbf{4.1}&\textbf{78.7}&\textbf{21.3}\\\hline
\end{tabular}
\newpage
\footnotesize\textbf{Table.7}   The features of the UNSW-NB15 dataset (taken by Ref.\cite{Ref31}) \\ 
\footnotesize\renewcommand{\arraystretch}{1}
\begin{tabular}{|c|c|c|}
\hline
\textbf{\#}&\textbf{Name}&\textbf{description}\\
\hline
1&scrip&Source IP address\\
2&sport&Source port number \\
3&dstip& Destination IP address \\
4&dsport& Destination port numner \\
\textbf{5}&\textbf{proto}&\textbf{Protocol type, such as TCP, UDP.}\\
6&state& The states and its dependent protocol e.g., CON.\\
7&dur&Row total duration\\
8&sbytes& Source to destination bytes.\\
9&dbytes& Destination to source bytes.\\
10&sttl&Source to destination time to live.\\
11&dttl&Destination to source time to live.\\
12&sloss&Source packets retransmitted or dropped.\\
13&dloss&Destination packets retransmitted or dropped.\\
\textbf{14}&\textbf{service}&\textbf{Such as http, ftp, smtp, ssh, dns and ftp-data.}\\
15&sload&Source bits per second.\\
16&dload&Destination bits per second.\\
17&spkts&Source to destination packet count.\\
18&dpkts&Destination to source packet count.\\
19&swin&Source TCP window advertisement value.\\
20&dwin&Destination TCP window advertisement value.\\
21&Stcpb&Source TCP base sequence number.\\
22&dtcpb&Destination TCP base sequence number.\\
23&smeansz&Mean of the packet size transmitted by the srcip.\\
24&dmeansz&Mean of the packet size transmitted by the dstip.\\
25&trans{\_}depth&The connection of http request/response transaction.\\
26&res{\_}bdy{\_}len&The content size of the data transferred from http.\\
27&sjit&Source jitter.\\
28&djit&Destination jitter.\\
29&stime&Row start time.\\
30&ltime&Row last time.\\
31&sintpkt&Source inter-packet arrival time.\\
32&dintpkt&Destination inter-packet arrival time.\\
33&tcprtt&Setup round-trip time, the sum of 'synack' and 'ackdat'.\\
34&synack&The time between the SYN and the SYN{\_}ACK packets.\\
35&ackdat&The time between the SYN{\_}ACK and the ACK packets.\\
36&is{\_}sm{\_}ips{\_}ports&If srcip (1) = dstip (3) and sport (2) = dsport (4), assign 1 else 0.\\
37&ct{\_}state{\_}ttl&No. of each state (6) according to values of sttl (10) and dttl (11).\\
38&ct{\_}flw{\_}http{\_}mthd&No. of methods such as Get and Post in http service.\\
39&is{\_}ftp{\_}login&If the ftp session is accessed by user and password then  else 0. \\
40&ct{\_}ftp{\_}cmd&No of flows that has a command in ftp session.\\
41&ct{\_}srv{\_}src&No. of rows of the same service (14) and srcip (1) in 100 rows.\\
42&ct{\_}srv{\_}dst&No. of rows of the same service (14) and dstip (3) in 100 rows.\\
43&ct{\_}dst{\_}ltm&No. of rows of the same dstip (3) in 100 rows.\\
44&ct{\_}src{\_}ltm&No. of rows of the srcip (1) in 100 rows.\\
45&ct{\_}src{\_}dport{\_}ltm&No of rows of the same srcip (1) and the dsport (4) in 100 rows.\\
46&ct{\_}dst{\_}sport{\_}ltm&No of rows of the same dstip (3) and the sport (2) in 100 rows.\\
47&ct{\_}dst{\_}src{\_}ltm&No of rows of the same srcip (1) and the dstip (3) in 100 rows. \\
48&Attack{\_}cat&The name of each attack category.\\
49&Label&0 for normal and 1 for attack records\\ \hline
\end{tabular}
\end{center}

\begin{thebibliography}{}
%
%
\bibitem{Ref55}
Abualigah, Laith Mohammad, Ahamad Tajudin Khader, Essam Said Hanandeh, and Amir H. Gandomi. "A novel hybridization strategy for krill herd algorithm applied to clustering techniques." Applied Soft Computing 60 (2017): 423-435. doi:\url{https://doi.org/10.1016/j.asoc.2017.06.059}
\bibitem{Ref56}
Abualigah, Laith Mohammad, Ahamad Tajudin Khader, and Essam Said Hanandeh. "A combination of objective functions and hybrid Krill herd algorithm for text document clustering analysis." Engineering Applications of Artificial Intelligence 73 (2018): 111-125. doi: \url{https://doi.org/10.1016/j.engappai.2018.05.003}
\bibitem{Ref57}
Abualigah, Laith Mohammad, Ahamad Tajudin Khader, and Essam Said Hanandeh. "Hybrid clustering analysis using improved krill herd algorithm." Applied Intelligence 48, no. 11 (2018): 4047-4071. doi: \url{https://doi.org/10.1007/s10489-018-1190-6}
\bibitem{Ref58}
Abualigah, Laith Mohammad Qasim. Feature selection and enhanced krill herd algorithm for text document clustering. Berlin: Springer, 2019. doi: \url{https://doi.org/10.1007/978-3-030-10674-4}
\bibitem{Ref59}
Abualigah, Laith Mohammad, and Ahamad Tajudin Khader. "Unsupervised text feature selection technique based on hybrid particle swarm optimization algorithm with genetic operators for the text clustering." The Journal of Supercomputing 73, no. 11 (2017): 4773-4795. doi: \url{https://doi.org/10.1007/s11227-017-2046-2}
\bibitem{Ref60}
Abualigah, Laith Mohammad, Ahamad Tajudin Khader, and Essam Said Hanandeh. "A new feature selection method to improve the document clustering using particle swarm optimization algorithm." Journal of Computational Science 25 (2018): 456-466. doi: \url{https://doi.org/10.1016/j.jocs.2017.07.018}
\bibitem{Ref61}
Deng, Wu, Junjie Xu, and Huimin Zhao. "An improved ant colony optimization algorithm based on hybrid strategies for scheduling problem." IEEE access 7 (2019): 20281-20292. doi: \url{https://doi.org/10.1109/ACCESS.2019.2897580}
\bibitem{Ref62}
Deng, Wu, Huimin Zhao, Li Zou, Guangyu Li, Xinhua Yang, and Daqing Wu. "A novel collaborative optimization algorithm in solving complex optimization problems." Soft Computing 21, no. 15 (2017): 4387-4398. doi: \url{https://doi.org/10.1007/s00500-016-2071-8}
\bibitem{Ref63}
Deng, Wu, Huimin Zhao, Xinhua Yang, Juxia Xiong, Meng Sun, and Bo Li. "Study on an improved adaptive PSO algorithm for solving multi-objective gate assignment." Applied Soft Computing 59 (2017): 288-302. doi: \url{https://doi.org/10.1016/j.asoc.2017.06.004}
\bibitem{Ref68}
Arqub, Omar Abu, and Zaer Abo-Hammour. "Numerical solution of systems of second-order boundary value problems using continuous genetic algorithm." Information sciences 279 (2014): 396-415. doi: \url{https://doi.org/10.1016/j.ins.2014.03.128}
\bibitem{Ref70}
Arqub, Omar Abu, AL-Smadi Mohammed, Shaher Momani, and Tasawar Hayat. "Numerical solutions of fuzzy differential equations using reproducing kernel Hilbert space method." Soft Computing 20, no. 8 (2016): 3283-3302. doi: \url{https://doi.org/10.1007/s00500-015-1707-4}
\bibitem{Ref64}
Rizvi, Baqar, Ammar Belatreche, and Ahmed Bouridane. "Immune Inspired Dendritic Cell Algorithm for Stock Price Manipulation Detection." In Proceedings of SAI Intelligent Systems Conference, pp. 352-361. Springer, Cham, 2019. doi: \url{https://doi.org/10.1007/978-3-030-29516-5_27}
\bibitem{Ref65}
Almasalmeh, Nuha, Firas Saidi, and Zouheir Trabelsi. "A Dendritic Cell Algorithm Based Approach for Malicious TCP Port Scanning Detection." In 2019 15th International Wireless Communications {\&} Mobile Computing Conference (IWCMC), pp. 877-882. IEEE, 2019. doi: \url{https://doi.org/10.1109/IWCMC.2019.8766461}
\bibitem{Ref66}
Elisa, Noe, Longzhi Yang, and Fei Chao. "Signal Categorisation for Dendritic Cell Algorithm Using GA with Partial Shuffle Mutation." In UK Workshop on Computational Intelligence, pp. 529-540. Springer, Cham, 2019. doi: \url{https://doi.org/10.1007/978-3-030-29933-0_44}
\bibitem{Ref67}
Nnko, Noe, Longzhi Yang, Xin Fu, and Nitin Naik. "Dendritic Cell Algorithm Enhancement Using Fuzzy Inference System for Network Intrusion Detection." (2019): 1-6. url: \url{http://nrl.northumbria.ac.uk/id/eprint/38822}
\bibitem{Ref69}
Hatamlou, Abdolreza. "Black hole: A new heuristic optimization approach for data clustering." Information sciences 222 (2013): 175-184. doi: \url{https://doi.org/10.1016/j.ins.2012.08.023}
\bibitem{Ref3}
P. Wlodarczak, Cyber immunity a bio-inspired cyber defense system (2017)
199–208 doi:\url{10.1007/978-3-319-56154-7_19.}
\bibitem{Ref12}
M. Pavone, G. Narzisi, G. Nicosia, Clonal selection: an immunological
algorithm for global optimization over continuous spaces, Journal of Global
Optimization 53 (4) (2012) 769–808. doi:\url{10.1007/s10898-011-9736-8.}
\bibitem{Ref15}
J. L. Santanelli, F. B. de Lima Neto, Network intrusion detection using danger theory
and genetic algorithms (2016) 394-405 doi:\url{10.1007/978-3-319-53480-0_39.}
\bibitem{Ref16}
Z. Chelly, Z. Elouedi, A survey of the dendritic cell algorithm, Knowledge and Information Systems 48 (3) (2016) 505–535. doi:\url{10.1007/s10115-015-0891-y.}
\bibitem{Ref17}
J. Greensmith, U. Aickelin, The deterministic dendritic cell algorithm, in:
International Conference on Artificial Immune Systems, Springer, 2010, pp.
291–302. doi:\url{10.1007/978-3-540-85072-4_26.}
\bibitem{Ref19}
D. Dal, S. Abraham, A. Abraham, S. Sanyal, M. Sanglikar, Evolution
induced secondary immunity: An artificial immune system based intrusion detection system, in: Computer Information Systems and Industrial Management Applications, 2008. CISIM’08. 7th, IEEE, 2008, pp. 65–70. doi:\url{10.1109/CISIM.2008.31.}
\bibitem{Ref28}
Greensmith, The dendritic cell algorithm, Ph.D. thesis, Citeseer (2007).
\bibitem{Ref31}
N. Moustafa, J. Slay, The significant features of the unsw-nb15 and
the kdd99 data sets for network intrusion detection systems, in: Building Analysis Datasets a nd Gathering Experience Returns for Security(BADGERS), 2015 4th International Workshop on, IEEE, 2015, pp. 25–31.
doi:\url{10.1109/BADGERS.2015.014.}
\bibitem{Ref32}
N. B. I. Azuan Ahmad, M. N. Kama, Cloudids: Cloud intrusion detec-
tion model inspired by dendritic cell mechanism, International Journal of
Communication Networks and Information Security (IJCNIS) Vol 9 (2017)
67–75.
\bibitem{Ref34}
B. Setiawan, S. Djanali, T. Ahmad, A study on intrusion detection using
centroid-based classification, Procedia Computer Science 124 (2017) 672–6
81. doi:\url{10.1016/j.procs.2017.12.204.}
\bibitem{Ref36}
G. Gu, P. Fogla, D. Dagon, W. Lee, B. Skoric, Measuring intrusion detection capability: an information theoretic approach (2006) 90–101 doi: \url{10.1145/1128817.1128834.}
\bibitem{Ref37}
O. Z. Eesa A.S, B. A.M.A., A novel feature-selection approach based on the
cuttlefish optimization algorithm for intrusion detection systems, Expert Systems with Applications 42 (5) (2015) 2670–2679. doi:\url{10.1016/j.eswa.2014.11.009.}
\bibitem{Ref44}
Dagdia, Zaineb Chelly. "A scalable and distributed dendritic cell algorithm for big data classification." Swarm and Evolutionary Computation (2018). doi: \url{https://doi.org/10.1016/j.swevo.2018.08.009}
\bibitem{Ref46}
Elisa, Noe, Longzhi Yang, and Nitin Naik. "Dendritic cell algorithm with optimised parameters using genetic algorithm." In 2018 IEEE Congress on Evolutionary Computation (CEC), pp. 1-8. IEEE, 2018. doi: \url{https://doi.org/10.1109/CEC.2018.8477932}
\bibitem{Ref33}
F. Gu, Theoretical and empirical extensions of the dendritic cell algorithm, Ph.D. thesis,
University of Nottingham (2011).
\bibitem{Ref73}
Elisa, Noe, Jie Li, Zheming Zuo, and Longzhi Yang. "Dendritic cell algorithm with fuzzy inference system for input signal generation." In UK workshop on computational intelligence, pp. 203-214. Springer, Cham, 2018. doi: \url{https://doi.org/10.1007/978-3-319-97982-3_17}
\bibitem{Ref74}
Elisa, Noe, Longzhi Yang, Yanpeng Qu, and Fei Chao. "A revised dendritic cell algorithm using k-means clustering." In 2018 IEEE 20th International Conference on High Performance Computing and Communications; IEEE 16th International Conference on Smart City; IEEE 4th International Conference on Data Science and Systems (HPCC/SmartCity/DSS), pp. 1547-1554. IEEE, 2018. doi: \url{https://doi.org/10.1109/HPCC/SmartCity/DSS.2018.00254}
\bibitem{Ref75}
Dagdia, Zaineb Chelly. "A distributed dendritic cell algorithm for big data." In Proceedings of the Genetic and Evolutionary Computation Conference Companion, pp. 103-104. ACM, 2018. doi: \url{https://doi.org/10.1145/3205651.3205701}
\bibitem{Ref76}
Zhao, Huimin, Rui Yao, Ling Xu, Yu Yuan, Guangyu Li, and Wu Deng. "Study on a novel fault damage degree identification method using high-order differential mathematical morphology gradient spectrum entropy." Entropy 20, no. 9 (2018): 682. doi: \url{https://doi.org/10.3390/e20090682}
\bibitem{Ref77}
Zhao, Huimin, Jianjie Zheng, Junjie Xu, and Wu Deng. "Fault diagnosis method based on principal component analysis and broad learning system." IEEE Access 7 (2019): 99263-99272. doi: \url{https://doi.org/10.1109/ACCESS.2019.2929094}
\end{thebibliography}
\end{document}